\UseRawInputEncoding
\documentclass[12pt]{iopart}

\usepackage{iopams}  
\usepackage{bm,amstext,graphicx,amssymb}  
\usepackage{color}

\begin{document}
\title{Untwining multiple parameters at the exclusive zero-coincidence points with quantum control}

\author{Yu Yang$^{1,*}$ \& Federico Belliardo$^2$ \& Vittorio Giovannetti$^2$ \& Fuli Li$^{1,*}$}
\address{$^1$Ministry of Education Key Laboratory for Nonequilibrium Synthesis and Modulation of Condensed Matter, Shaanxi Province Key Laboratory of Quantum Information and Quantum Optoelectronic Devices, School of Physics, Xi’an Jiaotong University, Xi’an 710049, China}
\address{$^2$NEST, Scuola Normale Superiore and Istituto Nanoscienze-CNR, I-56127 Pisa, Italy}
\eads{\mailto{yangyu1229@hotmail.com}, \mailto{flli@xjtu.edu.cn}}
\vspace{10pt}

\begin{abstract}
In this paper we address a special case of ``sloppy" quantum estimation procedures which happens in the presence of {\it intertwined} parameters. A collection of parameters are said to be intertwined when their imprinting on the quantum probe that mediates the estimation procedure, is performed by  a set of linearly dependent generators. Under this circumstance the  individual values of the parameters can not be recovered unless one tampers with the encoding process itself. An example  is presented by studying the estimation of the relative time-delays that accumulate along two parallel optical transmission lines. In this case  we show that the parameters can be effectively untwined by inserting a sequence of balanced beam splitters (and eventually adding an extra phase shift on one of the lines) that couples the two lines at regular intervals in a setup that  remind us a  generalized Hong-Ou-Mandel (GHOM) interferometer. For the case of two time delays we prove that, when the employed probe is the frequency-correlated biphoton state, the untwining occurs in correspondence of exclusive zero-coincidence (EZC) point. Furthermore we show the statistical independence of two time delays and the optimality of the quantum Fisher information at the EZC point. Finally we prove the compatibility of this scheme by checking the weak commutativity condition associated with the symmetric logarithmic derivative operators.
\end{abstract}

%
\vspace{2pc}
\noindent{\it Keywords}: quantum multiparameter estimation, quantum Fisher information matrix, generalized Hong-Ou-Mandel interferometry 

%
\submitto{\NJP}
%
\maketitle
%
%

\section{Introduction}
Quantum metrology~\cite{QET_book,holevo2011probabilistic,QM} targets the problem of estimating physical quantities as precise as possible by exploiting the advantages offered by quantum coherence and (unconventional) quantum measurements, so as to obtain a higher estimation precision than the classical estimation scheme.
Compared to the extensively investigated quantum estimation of a single parameter, the quantum estimation of multiple parameters is more important for many practical applications such as the magnetometry~\cite{PhysRevLett.116.030801},  gyroscopy~\cite{PhysRevA.95.012326}, or quantum network~\cite{PhysRevLett.120.080501}. 
However, the much challenges the quantum multi-parameter estimation task has, since it is extremly difficult to simultaneously and optimally recover a collection of (say) $k$ independent (unknown) parameters from the same experimental setup ~\cite{Multi_1,Multi_2,Multi_3,tight_bound,Incom_measurement,multiReview,Mu_satu3,Review}.
The most general multi-parameter estimation task can be casted in a black-box scenario~\cite{qmetrology} where
the collection of to-be-estimated quantities  expressed by the real vector ${\bm \tau}=\{\tau_1,\cdots, \tau_k\}$ are initially imprinted  into the state of a quantum probing system via an encoding map $\Lambda_{\bm \tau}$ that depends parametrically upon $\bm{\tau}$.
The value of $\bm \tau$  is hence recovered through measurements  performed on a statistically significant set of $M$ copies of the imprinted state of the probe. Under these conditions the ultimate precision of  multiple parameters is gauged by  the Quantum Cram\'er Rao bound (QCRB)~\cite{QFI_Origin,QFI_Origin2,QFI} that establishes a lower bound for 
Mean Square Error {(MSE)} matrix of the problem via the  inverse of the Quantum Fisher Information Matrix (QFIM).
It is well known however that in many cases of physical interests the multi-parameter QCRB precision threshold cannot be attained (not even in the asymptotic limit of large $M$)  due to the compatibility problems pertaining the non-commutativity of quantum operations~\cite{tight_bound,Incom_measurement,Mu_satu3,Heinosaari_2016,Mu_satu,Mu_satu2}.
An even more drastic limitation occurs when one faces what we may call an {\it intertwined} multi-parameter scenario, i.e. when the black-box map $\Lambda_{\bm \tau}$ admits the same generator for some of the $k$ components of the vector ${\bm \tau}$ or, more generally, when a subset of such parameters have linearly dependent generators. 
When this happens the QFIM turns out to be singular, the QCRB to be meaningless and 
the individual components of ${\bm \tau}$  cannot be individually discerned and estimated~{\cite{TAM}}.
Examples of this behavior are well known in computation biology and chemistry where they are typically identified as {\it sloppy models} \cite{GUTE,TRANS,MANNA}.
It is clear that in these cases neither a careful optimization of the input state of the probe $\hat{\rho}$, nor a careful choice of the measurement procedure will enable us to recover the values of ${\bm \tau}$.

In the present manuscript we present an example of this phenomenon considering the estimation of multiple time delays $\tau_1, \tau_2, \cdots, \tau_k$ on a transmission line, with respect to a second reference line, which could be generated for example by length differences in the paths. It has been shown that the estimation precision with respect to the time delay can approach the attosecond scale (i.e. the length-difference reaches to the nanometer scale)~\cite{HOM_Past1}. 
However, by sending light on this couple of lines and measuring it at the end we can only estimate the total time delay (or the mean delay for unit of length in the transmission line), but the information on the spatial distributions of such delays is totally inaccessible to us, because it hasn't been codified in a discernible way in the first place.
One motivation of our study is exactly to explore a way to simultaneously estimate (or untwine) these intertwined parameters with the individual highest precision, i.e. to achieve the simultaneous optimal estimation of multiple intertwined parameters.
The only possibility we find at our disposal is to interfere directly with the encoding process $\Lambda_{\bm \tau}$, e.g. by means of quantum control acting between the encoding of the different parameters. 
Clearly a possible solution is to open the transmission line at regular time interval to read each individual delay. While effective, this strategy has however the major drawback that it requires us to effectively destroy the transmission line. In contrast to this active and invasive technique we show that the passive solution exists that can do the same job without signal leakage.
Concretely, the solution allowing us to effectively {\it untwining}  the imprinting of the various parameters,  is to alternate the encoding of the individual delays with a unitary control of the probe, so that each parameter will be encoded in a different way and can be discerned from the others by a measurement at the end of the lines. 
The simplest form of unitary control for light traveling on a couple of transmission lines is a 50:50 beam splitter (BS). If we insert a BS between each encoding of the time-delay, and add an achromatic phase~\cite{ACHRO3}, to further generalize the unitary control we are led automatically to the setting of the generalized Hong-Ou-Mandel (GHOM) interferometer, which was discussed in reference~\cite{GHOM}. The GHOM interferometer is an extension of the ``HOM effect"~\cite{HOM} that exhibits what was dubbed exclusive zero-coincidence (EZC) points, i.e. a one-to-one correspondence between the contemporary absence of all the time delays and zero values of the coincidence counts at the output of the device when the input state of the device is a frequency-correlated  biphoton state. 
Therefore, we take the GHOM setting as a representative scheme of untwining multiple interwined parameters by introducing quantum control specified as the 50:50 BS and achromatic phase.
We then compute the QFI for a two-photon symmetric input state at the EZC point, explicitly obtaining an invertible matrix, and proving therefore the success of the untwining of the time delays. We observe also that at the EZC point the QFIM looses the off-diagonal components (hence ensuring the statistical independence of the estimation of $\tau_1$ and $\tau_2$), while attaining the maximum values of the diagonal entries (therefore providing the maximal estimation accuracy for both the parameters).

The manuscript is organized as follows: in section~\ref{Sec:QETR} we give a brief review of the theory of quantum parameter estimation, formalize the {intertwined} multi-parameter scenario, and discuss
the compatibility issues that  may arise  when the parameters are not {intertwined}.
In section~\ref{SECIII} we present out {a} case study based on  the estimation of multiple time delays 
along an optical transmission line. Here after recalling some basic facts about the GHOM interferometer we show how for the case of {two} delays, this setup can be used to effectively {untwine such parameters}. We also show that at the EZC point the time-delay parameters are also statistically independent and that the  QCRB can be saturated.
In the last part of section~3, we also discuss the singularities of the different QFIMs in the cases of three or four time delays. Some conclusions and the possible applications are presented  in section~\ref{Sec:Conclusion}.

\section{Quantum multi-parameter estimation}\label{Sec:QETR}
In this section we review some basic mathematical tools that, on one hand, allow us to  identify those cases where
the multi-parameter estimation is formally impossible (intertwined configurations), and on the other hand permit to
clarify the compatibility issues that affect the efficiency of those where the estimation is possible. 
\subsection{Intertwined vs not-intertwined configurations with quantum control} \label{DEFintert} 
Consider a multi-parameter quantum estimation scenario in which a  collection of $k$ unknown parameters represented by the vector $\bm{\tau}=\{\tau_1,\cdots, \tau_k\}$ are imprinted on a probing quantum system via an assigned mapping 
\begin{eqnarray} \label{MAPPA} 
	\hat{\rho}_{\bm \tau} =\Lambda_{\bm \tau} (\hat{\rho}) \;,
\end{eqnarray}
($\hat{\rho}$ being the input state of the probe). A characterization of  the   precision attainable in the process is provided by the $k\times k$, positive semidefinite, QFIM $\bm{\mathcal{H}_{\tau}}$ of the model~\cite{QFI_Origin,QFI_Origin2,QFI,Do1,Do2} whose elements can be expressed via the spectral decomposition $\hat{\rho}_{\bm \tau}=\sum_l p_l |l \rangle\langle l |$  of the probe state as 
\begin{eqnarray}\label{eq:QFIM}
	[\bm{\mathcal{H}_{\tau}}]_{ij}:=2\sum_{p_l+p_{l'} \neq 0}\frac{\langle {l'}|\partial_i \hat{\rho}_{\bm \tau}|l\rangle \langle l|\partial_j \hat{\rho}_{\bm \tau}|{l'}\rangle}{p_l+p_{l'}}\;,
\end{eqnarray}
an identity which for {the pure state} $\hat{\rho}_{\bm \tau}=|\Psi_{\bm \tau}\rangle \langle \Psi_{\bm \tau}|$ simplifies into
\begin{eqnarray}\label{eq:QFIM_new}
	[\bm{\mathcal{H}_{\tau}}]_{ij}=4\text{Re}[\langle \partial_i \Psi_{\bm \tau}|\partial_j \Psi_{\bm \tau}\rangle-\langle \partial_i \Psi_{\bm \tau} | \Psi_{\bm \tau}\rangle \langle \Psi_{\bm \tau}|\partial_j \Psi_{\bm \tau} \rangle],
\end{eqnarray}
(in the above expressions $\partial_\ell$ represents the partial derivative with respect to $\tau_\ell$, while $\text{Re}[\bullet]$ means extracting the real part). 

We can now distinguish two different scenarios depending on the invertibility of the QFIM.
If  $\bm{\mathcal{H}_{\tau}}$ is invertible (i.e. if $\mbox{Det}[\bm{\mathcal{H}_{\tau}}] > 0$)  for at least one special choice of the input state $\hat{\rho}$,  then the recovering of the values of $\bm \tau$ is possible (at least in principle) with an  ultimate estimation precision   that is gauged by the QCRB~\cite{QFI_Origin,QFI_Origin2,QFI,Do1,Do2}, i.e.  
\begin{eqnarray}\label{Multi_QCRB}
	\mathbf{Cov}[\tilde{{\bm \tau}}] \ge \frac{1}{M} \bm{\mathcal{H}_{\tau}}^{-1}\;,
\end{eqnarray}
where $M$ is the number of copies of $\hat{\rho}_{\bm \tau}$ we have access to, and where $\mathbf{Cov}[\tilde{{\bm \tau}}]$  is the covariance matrix of elements $[\mathbf{Cov}[\tilde{{\bm \tau}}]]_{ij}:=\text{E}[\tilde{\tau}_i \tilde{\tau}_j]-\text{E}[\tilde{\tau}_i] \text{E}[\tilde{\tau}_j]$ with $\tilde{{\bm \tau}}:=\{\tilde{\tau}_1,\cdots,\tilde{\tau}_k\}$ being the estimated values of the vector ${{\bm \tau}}$.
The positivity of $\bm{\mathcal{H}_{\tau}}$ implies $[\bm{\mathcal{H}_{\tau}}^{-1}]_{ii} \ge 1/[\bm{\mathcal{H}_{\tau}}]_{ii}$ for any $i$, that in turns results in the following inequality~\cite{Review}
\begin{eqnarray}\label{tmp1}
		\Delta^2 {\tau}_i \ge \frac{1}{M} [\bm{\mathcal{H}_{\tau}}^{-1}]_{ii} \ge \frac{1}{M [\bm{\mathcal{H}_{\tau}}]_{ii}}\;,
\end{eqnarray}
the identity between the second and third term being achieved when $\bm{\mathcal{H}_{\tau}}$ is diagonal, ensuring the statistical independence of the estimation of the individual parameters of the model. Equation~(\ref{tmp1}) proves the consistency of equation~(\ref{Multi_QCRB}) with the bound one would obtain in the special case where one attempts to recover the $i$-th component of ${\bm \tau}$ in the scenario where the remaining $k-1$ components of such vector are known.

\begin{figure*}[!t]
	\centering
	\includegraphics[width=0.7\textwidth]{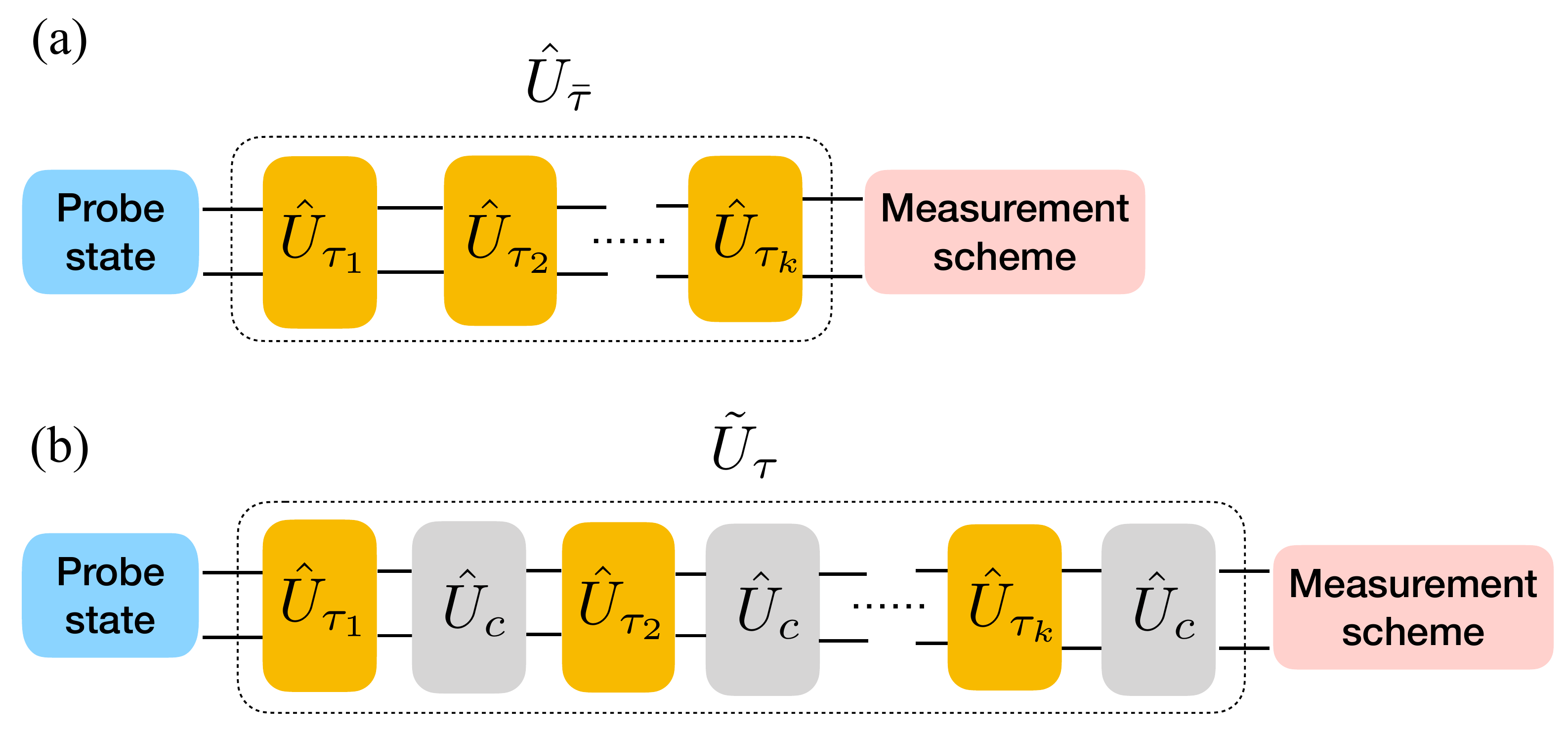}
	\caption{Panel (a) gives a paradigmatic scheme of the intertwined multi-parameter scenario, in which the unknown parameters $\{\tau_1,\tau_2,\cdots,\tau_k\}$ are encoded by a sequence of unitary transformations $\hat{U}_{\tau_1}$, $\hat{U}_{\tau_2}$, $\cdots$, $\hat{U}_{\tau_k}$ and they compose a whole unitary transformation $\hat{U}_{\bar \tau}$. Panel (b) gives an optimized scheme of panel (a), in which quantum control $\hat{U}_c$ acts between the encoding of the different parameters, the whole unitary transformation is described by $\tilde{U}_{\bm \tau}$.}
	\label{FigScheme}
\end{figure*}

If on the contrary $\bm{\mathcal{H}_{\tau}}$ is not invertible (i.e. if $\mbox{Det}[\bm{\mathcal{H}_{\tau}}] = 0$) for all possible choices of the input state of the probe, we are in presence of an intertwined multi-parameter estimation scenario in which the mapping~(\ref{MAPPA}) is characterized by linearly dependent generators for each of the $k$ components  $\tau_1$, $\cdots$, $\tau_k$. A paradigmatic example of this sort  occurs e.g. when $\Lambda_{\bm \tau}$ is induced by a sequence of unitary transformations $\hat{U}_{\tau_1}$, $\hat{U}_{\tau_2}$, $\cdots$, $\hat{U}_{\tau_k}$ generated by the same Hamiltonian generator $\hat{H}$ i.e. 
\begin{eqnarray} 
	\hat{U}_{\tau_j} &=&\exp{(}- i \hat{H} \tau_j {)}\;, \qquad \forall j=1,\cdots, k\;,
	\\
	\Lambda_{\bm \tau}{(}\hat{\rho}{)} &=& \hat{U}_{\tau_k} \cdots \hat{U}_{\tau_1} \hat{\rho} 
	\hat{U}^\dag_{\tau_1} \cdots \hat{U}^\dag_{\tau_k}= \hat{U}_{\overline{\tau}} \hat{\rho}  \hat{U}^\dag_{\overline{\tau}}\;,   \label{defoverline}   
\end{eqnarray} 
with $\overline{\tau} := \sum_{j=1}^k \tau_j$ (an explicit instance of this kind of model is given in the next section). 
Clearly under these special circumstances the probe state  $\hat{\rho}_{\bm \tau}$, irrespectively from the choice of its input configuration $\hat{\rho}$,  will only carry information  on the linear combination $\overline{\tau}$ making the reconstruction of the individual components of $\bm \tau$ impossible (a fact signalled by the loss of meaning of (\ref{Multi_QCRB}) whose right-hand-side is now effectively divergent). 
In other words, in these cases neither a careful optimization of the initial probe state $\hat{\rho}$, nor a careful choice of the measurement procedure will enable us to untwine ${\bm \tau}$. 
In the face of this difficulty, we can resort to control-enhanced quantum parameter estimation procedures, i.e. interfering with the probe evolution between the encoding of the different parameters.
References~\cite{Control1,Control2} already demonstrated that with the help of quantum controls acting the adjacent two systems in a sequence structure, the optimized quantum dynamics is capable to achieve the multi-parameter simultaneous optimal estimation.
Accordingly it is significant to explore  the possibility of untwining the multiple parameters by alternating the encoding of individual parameters with quantum controls, as shown in figure~\ref{FigScheme} (b). In this case equation~(\ref{defoverline}) is renewed as
\begin{eqnarray}\label{eq:control}
	\tilde{\Lambda}_{\bm \tau}{(}\hat{\rho}{)} =\hat{U}_c \hat{U}_{\tau_k} \cdots \hat{U}_c \hat{U}_{\tau_1} \hat{\rho} \hat{U}^\dag_{\tau_1}\hat{U}_c^\dag \cdots \hat{U}^\dag_{\tau_k}\hat{U}_c^\dag= \tilde{U}_{\bm \tau} \hat{\rho}  \tilde{U}^\dag_{\bm \tau}\;,
\end{eqnarray} 
where $\hat{U}_c$ is the introduced quantum control and $\tilde{U}_{\bm \tau}$ plays a same role as $\hat{U}_{\overline{\tau}} $ in equation~(\ref{defoverline}). It is necessary to stress that while certainly $\tilde{U}_{\bm{\tau}} \neq \hat{U}_{\bar{\tau}}$, the two schemes are still produced by the same local encoding steps $\hat{U}_{\tau_k}$'s: they only differ in the way such local steps are allowed to operate on the system. In the following section~\ref{SECIII}, we take the GHOM interferometry as an explicit example to clarify the feasibility of untwining multiple delays.
\subsection{Compatibility and asymptotic achievability of the QCRB} 
A second main issue associated with multi-parameter estimation procedures is that the imperfect knowledge of one of the parameters tends to deteriorate the precision of estimating the others. This implies that for $k\geq 2$ the inequality~(\ref{Multi_QCRB}) is typically not reachable even in the asymptotic limit of large $M$ values. Sufficient requirements for this to happen have been identified~\cite{multiReview,Mu_satu3,Mu_satu,Mu_satu2} when certain commutativity conditions hold true for the symmetric logarithmic derivative (SLD) operator
\begin{eqnarray}
	\label{SLDdef} \hat{L}_j:=2 \sum\limits_{p_l+p_{l'} \neq 0} \frac{\langle l|\partial_j \hat{\rho}_{\bm \tau}|l'\rangle}{p_l+p_l'}|l\rangle \langle l'|\;,
\end{eqnarray} 
which allows one to express the variation of $\hat{\rho}_{\bm \tau}$ {with} the $j$-th component of the vector ${\bm \tau}$ in terms of the following Lyapunov matrix equation $\partial_j \hat{\rho}_{\bm \tau}=({\hat{L}_j \hat{\rho}_{\bm \tau}+\hat{\rho}_{\bm \tau} \hat{L}_j})/{2}$. Specifically we can say that the inequality (\ref{Multi_QCRB}) saturates for sufficiently large $M$ if does {happen} that the following condition holds true
\begin{eqnarray}\label{weak_comm}
	\Tr[\hat{\rho}_{{\bm \tau}}[\hat{L}_i,\hat{L}_j]]=0\;, \qquad \forall i,j\in\{ 1, \cdots, k\}\;, 
\end{eqnarray}
an identity that for {the pure state} $\hat{\rho}_{\bm \tau}=|\phi_{\bm \tau}\rangle \langle \phi_{\bm \tau}|$ {simplifies} into
\begin{eqnarray}\label{weakcomm_new}
	\text{Im}[\langle \partial_i \Psi_{\bm \tau}|\partial_j \Psi_{\bm \tau}\rangle]=0\;, \qquad \forall i,j\in\{ 1, \cdots, k\}\;, 
\end{eqnarray}
using the fact that in this case (\ref{SLDdef}) rewrites as~\cite{Review}
\begin{eqnarray}\label{weakcomm_new-1}
	\hat{L}_j=2(| \partial_j \Psi_{\bm \tau}\rangle \langle \Psi_{\bm \tau}|+| \Psi_{\bm \tau}\rangle \langle \partial_j \Psi_{\bm \tau}|)\;.
\end{eqnarray}
For a pure encoded state $|\Psi_{\bm \tau}\rangle$ the weak commutativity condition~(\ref{weakcomm_new}) implies also the strong commutativity, that is, there exists a set of SLD {operators} such that $[\hat{L}'_i, \hat{L}'_j] = 0$~\cite{multiReview}. When the encoded state is pure the SLD operators are not univocally specified, and if the weak commutativity holds, this freedom allows us to find a set of commuting SLD operators. The common {bases} of {these} operators {can be used to design} the projective measurement to be realized on $|\Psi_{\bm \tau}\rangle$ having the Classical Fisher Information Matrix (CFIM) reaching the QFIM. The QCRB can then be saturated asymptotically in $M$ with a maximum likelihood estimator for example.

Besides as discussed in section~\ref{DEFintert}, the behavior of the QFIM $\bm{\mathcal{H}_{\tau}}$ can be changed from non-invertible to invertible such that the intertwined parameters can be untwined by optimizing the original quantum dynamics with quantum controls. Accordingly, another problem is  whether these parameters can be recovered simultaneously and optimally,  i.e. investigating the compatibility of the scheme depicted by figure~\ref{FigScheme} (b) by checking the weak commutativity condition~(\ref{weak_comm}) associated with the SLD operators~(\ref{weakcomm_new-1}). In the following section~\ref{SECIII}, we take the GHOM interferometry as an explicit example to investigate the asymptotic achievability of the associated QCRB as a result of the mentioned compatibility.

\section{A case study} \label{SECIII} 
\begin{figure*}[!t]
	\centering
	\includegraphics[width=0.9\textwidth]{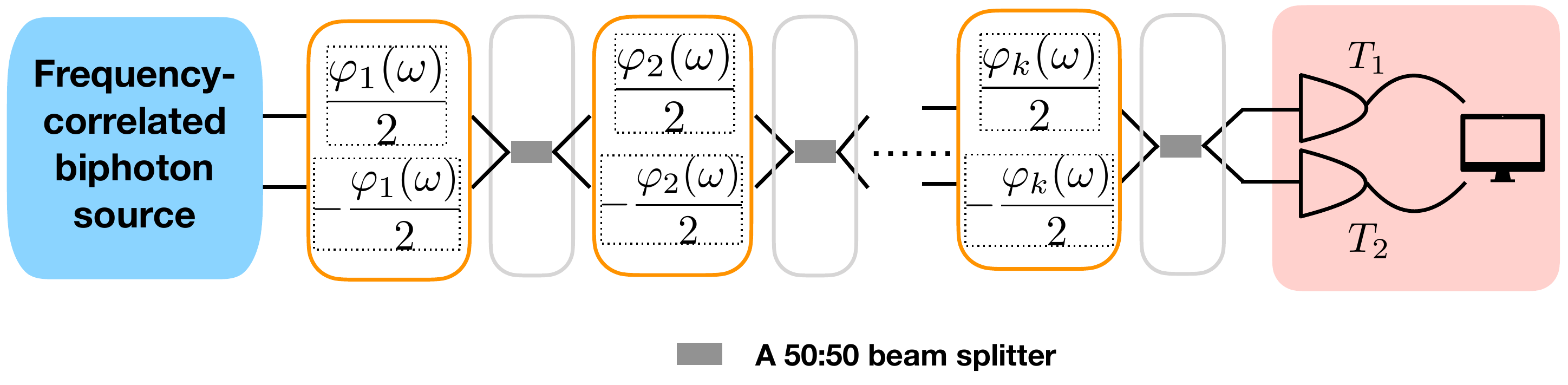}	
	\caption{ A sketch of the generalized Hong-Ou-Mandel (GHOM) interferometer, it includes $k$ cascaded phase-shift modules that induce two opposite phase shifts $\varphi_\ell(\omega)/2$ and $-\varphi_\ell(\omega)/2$ ($\ell=1,2,\cdots, k$) along the upper and lower arms respectively. The initial probe state is	produced from the frequency-correlated biphoton source and  two single-photon detectors $T_1$, $T_2$ are used for measuring. This GHOM setting can be used for untwining the delays and allows for the recovering of the individual delays, in which the balanced beamsplitters plus some possible achromatic phases play the role of quantum controls so as to achieve the control-enhanced multi-delays estimation.}
	\label{Fig:GHOM}
\end{figure*}

In this section we focus on an explicit example of an intertwined multi-parameter model based on {the} quantum optical setting. As shown in panel (a) of figure~\ref{FigScheme} it consists into  two parallel multimode optical lines connecting a sender to a remote receiver while experiencing $k$ relative temporal delays $\tau_1$, $\cdots$, $\tau_k$ that are spatially distributed along the path, which results into a global delay of the emergent signals represented exactly by the quantity~$\overline{\tau}$ of equation~(\ref{defoverline}). As discussed in section~\ref{DEFintert} under these circumstances there is no way that the two parties will be able to recover (even approximatively) the individual values of the $\tau_1$, $\cdots$, $\tau_k$. Of course a solution of the problem would be to grant access to each of the individual portions of the line where the delays are introduced: it is clear  however that this choice requires a complete redesign of the setup which will have a huge impact on the blue print of the original model (one communication line connects two remote parties). What we are looking for instead is a much less invasive  modification for the scheme. For this purpose we adopt the GHOM interferometric setup introduced in reference~\cite{GHOM} which only accounts for introducing sequences of balanced beam splitters  connecting the upper and lower lines of the scheme, plus possibly a collection of  achromatic phases~\cite{ACHRO3} -- see figure~\ref{Fig:GHOM}. In this GHOM setting quantum controls are specified as the balanced beamsplitters plus some achromatic phases so as to achieve the control-enhanced multi-delays estimation. As a matter of fact, the simplest form of unitary control for light traveling on a couple of transmission lines is exactly a 50:50 BS. As we shall see explicitly for the case of $k=2$, this choice is successful as the new QFIM of the scheme associated with a two-photon input state, is explicitly non-singular, hence allowing us to untwine two time delays. Interestingly enough the proposed setup also permits to fulfill the weak commutativity condition~(\ref{weakcomm_new}), hence ensuring the possibility of asymptotically reaching the associated QCRB~(\ref{Multi_QCRB}), and (for the special case of $\tau_1=\tau_2=0$) the saturation of the second inequality in (\ref{tmp1}) with maximum values for diagonal QFI components. 

\subsection{The GHOM interferometer} 
Here we briefly review the main feature of the GHOM interferometric setup introduced in reference~\cite{GHOM} as a method for {pin-pointing} the contemporary zero values of multiple independent time-delay parameters expressed by the vector $\bm{\tau}=\{\tau_1, \cdots, \tau_k\}$. Figure~\ref{Fig:GHOM} gives the schematic diagram of the setting, which includes $k$ cascaded phase-shift modules that induce  two opposite phase shifts $\varphi_\ell(\omega)/2$ and $-\varphi_\ell(\omega)/2$ ($\ell=1,2,\cdots, k$) along the upper and lower arms respectively. The $k$ uses of phase-shift modules are interleaved with $k$ balanced BSs. This kind of allocation of phases is usually known as the symmetric phase-shift~\cite{SP1,SP2,SP3,Sensiti}. For $\ell>1$ the phase shift $\varphi_\ell(\omega)$ is constituted by {a} time-delay element $\tau_\ell$ (the $\ell$-th parameter we need to estimate), and by a frequency-independent achromatic wave-plate $\theta_\ell$~\cite{ACHRO3} that instead represents {the} control {parameter} of the setup, i.e.
\begin{eqnarray}\label{eq:phase_shift}
	\varphi_\ell(\omega)=\omega \tau_\ell+\theta_\ell{,} \qquad  \forall \; \theta_\ell \in [0,2\pi)\;.
\end{eqnarray}
Yet the first phase-shift module only contains a time delay element, i.e. $\varphi_1(\omega)=\omega \tau_1$, which corresponds to a conventional HOM interferometer~\cite{HOM}. The initial state of the interferometer is assumed to be a symmetric, frequency-correlated biphoton pure state
\begin{eqnarray}\label{eq:input}
	|\Psi\rangle\!=\! \int d\omega \! \int d\omega' \Psi_s(\omega,\omega') \hat{a}_1^{\dagger}(\omega) \hat{a}_2^{\dagger}(\omega') {|\varnothing\rangle},
\end{eqnarray}
where $|\varnothing\rangle$ stands for the Fock vacuum state, where $\hat{a}_{1}^{\dagger}(\omega)$ and $\hat{a}_{2}^{\dagger}(\omega)$ are Bosonic creation operators that describe a photon of frequency $\omega$ that enters the device along the upper and lower input port respectively, and where finally $\Psi_s(\omega,\omega')$ is the biphoton joint spectral amplitude (JSA) following the normalization condition $\int d\omega \int d\omega' |\Psi_s(\omega,\omega')|^2=1$, and possessing the exchanging symmetry $\Psi_s(\omega,\omega')	=\Psi_s(\omega',\omega)$. The output state that emerges from the interferometer can now be expressed as the sum of two orthogonal terms, i.e.
\begin{eqnarray}\label{out}
	|\Psi_{\bm \tau}\rangle&=& |\Phi_{\bm \tau}\rangle+|\Upsilon_{\bm \tau}\rangle \;,
\end{eqnarray}
with the (not necessarily normalized) vectors $|\Phi_{\bm \tau}\rangle$ and $|\Upsilon_{\bm \tau}\rangle$ representing respectively events in which the two photons of the model emerges either on the same ports (biphoton bunching) or in distinct port (biphoton anti-bunching) -- see~\ref{app:state} for details. 
Accordingly the probability of {the} coincidence counts where each detector at the output of the device captures only one {photon}, corresponds to 
\begin{eqnarray}\label{RBP}
	R(\bm{\tau}) := \langle \Upsilon_{\bm \tau}|\Upsilon_{\bm \tau}\rangle\;.
\end{eqnarray}
The presence of {an} EZC point in the model emerges by observing that for special values of $k$ there exists an optimal choice $\bm{\bar{\theta}} := \{\bar{\theta}_2, \cdots, \bar{\theta}_k\}$ of the achromatic wave-plates vector $\bm{\theta}:=\{\theta_2,\cdots,\theta_k\}$ such that a one-to-one correspondence {relation} can be established between the zero point of the coincidence counts and the contemporary absence of all the time delays~\cite{GHOM,2HOM}, i.e.
\begin{eqnarray}\label{relation}
	R(\bm{\tau})|_{\bm{\theta}=\bm{\bar{\theta}}} =0 \qquad \Longleftrightarrow \qquad \tau_1=\cdots=\tau_k=0\;,
\end{eqnarray}
In particular in  reference~\cite{2HOM} it has been shown that this happens for $k=2$  with $\bar{\theta}_2= \pi/2$, for $k=4$ with $\bar{\theta}_3=\arccos(\cot\bar{\theta}_2 \cot\bar{\theta}_4)$ under the assumption that  $\sin\bar{\theta}_2 \sin \bar{\theta}_4 \neq 0$, while notably no solutions exist for $k=3$.
Equation~(\ref{relation}) shows that under EZC conditions (i.e. for ${\bm{\theta}=\bm{\bar{\theta}}}$) the GHOM provides a method to simultaneously pin-point the zero values of all the components of the ${\bm \tau}$ vector: this naturally suggests that the same setting could be used to improve the efficiency of the estimation of the delays. In the next section we  check this fact by focusing on the simplest (yet not trivial) case $k=2$.
In this scenario we shall see that indeed by setting $\theta_2=\bar{\theta}_2$ not only the QFIM measured on the output signal of the GHOM setup can be made invertible,  but also that both the inequalities of equation~(\ref{tmp1})  are locally achieved for  $\bm{{\tau}}=0$. 
We stress however that the optimal measurements  that would lead to the saturation of the QCRB would not be the simple photon-detections  associated with the EZC condition~(\ref{relation}) but,  as explained at the end of section~\ref{Sec:QETR}, projective measures derived from the SLD operators~(\ref{weakcomm_new-1}).

\subsection{Joint time-delay estimation via a GHOM interferometer}\label{Sec:GHOM}
First of all in section~\ref{sec:statInd} we show that the QFI matrix~(\ref{eq:QFIM_new}) of the output state~(\ref{out}) of a $k=2$ GHOM interferometer is {non-singular}, by explicitly computing its entries and plotting its determinant. As anticipated at the beginning of the section, this ensures the possibility of recovering both $\tau_1$ and $\tau_2$ {is at variance with} what's happening in the original setting of panel (a) of figure~\ref{FigScheme} which only allows for the estimation of $\overline{\tau} = \tau_1 + \tau_2$.  We  also show that enforcing the  EZC condition~(\ref{relation}),  at this special point one gets the saturation of the second inequality of equation~(\ref{tmp1}) with maximum values of the diagonal QFIM elements. In section~\ref{sec:weak} we prove that there exists a measurement at the end of the two lines reaching the {QFIM}, by proving the validity of the weak commutativity condition~(\ref{weakcomm_new-1}) for the case of $k=2$ time delays, this is sufficient to the pure encoded state~\cite{multiReview}. Finally in section~\ref{sec:higher}, we numerically investigate the GHOM settings with three or four time delays. The similar results have been obtained in the case of  four time delays, i.e. the GHOM interferometer of $k=4$, under the EZC condition  ${\bm{\theta}=\bm{\bar{\theta}}}$ (specifically we have numerically tested the model using $\theta_2=\pi/3, \theta_3=\arccos\left({1}/{\sqrt{3}}\right), \theta_4=\pi/4$), gives a nonsingular QFIM. Notably instead for $k=3$ time delays we haven't been able to find a configuration of $\bm{\theta}$ such that the QFIM is nonsingular, a fact that mimics the observation of reference~\cite{2HOM} that no EZC point can be found under this condition.

\subsubsection{Invertibility of the QFIM} \label{sec:statInd}
We start observing that the bunching and anti-bunching components of the output state (\ref{out}) $|\Psi_{\bm \tau}\rangle$ remain orthogonal even under differentiation,~i.e.
\begin{eqnarray} \label{NEWIDE} 
		\eqalign{
		\langle\partial_i \Phi_{\bm \tau}| \Upsilon_{\bm \tau}\rangle=\langle \Phi_{\bm \tau}| \partial_j \Upsilon_{\bm \tau}\rangle=0\;,\\
		\langle\partial_i \Phi_{\bm \tau}|\partial_j \Upsilon_{\bm \tau}\rangle=0\;, }
\end{eqnarray} 
for all $i$, $j$. Inserting hence equation~(\ref{out}) into equation~(\ref{eq:QFIM_new}) and invoking the decomposition~(\ref{coin1}) of \ref{app:state}  we can write
\begin{eqnarray}\label{eq:completeQFI}
\hspace{-2.5cm}[\bm{\mathcal{H}_{\tau}}]_{ij} &\!=\! &4\text{Re}[\langle \partial_i \Psi_{\bm \tau}|\partial_j \Psi_{\bm \tau}\rangle \!+\! \langle \partial_i \Psi_{\bm \tau}|\Psi_{\bm \tau}\rangle \langle \partial_j \Psi_{\bm \tau}|\Psi_{\bm \tau} \rangle]  \nonumber \\
\hspace{-2.5cm}&\!=\!&4\text{Re} \! \Big[\! \langle \partial_i \Phi_{\bm \tau}|\partial_j \Phi_{\bm \tau} \! \rangle \!+\! \langle \partial_i \Upsilon_{\bm \tau}|\partial_j \Upsilon_{\bm \tau} \! \rangle
\!+\! (\! \langle \partial_i \Phi_{\bm \tau}|\Phi_{\bm \tau}\! \rangle \!+\! \langle \partial_i \Upsilon_{\bm \tau}|\Upsilon_{\bm \tau}\! \rangle ) \! ( \! \langle \partial_j \Phi_{\bm \tau}|\Phi_{\bm \tau}\! \rangle \!+\! \langle \partial_j \Upsilon_{\bm \tau}|\Upsilon_{\bm \tau}\! \rangle )\! \Big]\nonumber \\ 
\hspace{-2.5cm}&\!=\!&4\text{Re}\Big\{ \! \int\! d\omega \! \int\! d\omega' |\Psi_s(\omega,\omega')|^2 \label{FINALFORQ} \Big[\partial_i \mathfrak{m}^*_{11}(\omega,\omega')\partial_j \mathfrak{m}_{11}(\omega,\omega')\!+\!\partial_i\mathfrak{m}^*_{22}(\omega,\omega')\partial_j \mathfrak{m}_{22}(\omega,\omega')\nonumber\\ 
\hspace{-2.5cm}&\!+\!&\partial_i \mathfrak{m}^*_{12}(\omega,\! \omega') \partial_j \mathfrak{m}_{12}(\omega,\! \omega') \!+\! \mathfrak{h}_i\mathfrak{h}_j\Big]\!\Big\}, 
\end{eqnarray}
where in the second identity we used~(\ref{NEWIDE}), and where we defined 
\begin{eqnarray}
 \mathfrak{h}_\ell \!:=\! &\int& \! d\omega \! \int\! d\omega' |\Psi_s(\omega,\omega')|^2   
 \left[ \partial_\ell \mathfrak{m}^*_{11}(\omega,\omega')\mathfrak{m}_{11}(\omega,\omega')\right. \nonumber\\
&+&\left. \partial_\ell\mathfrak{m}^*_{22}(\omega,\!\omega')\mathfrak{m}_{22}(\omega,\!\omega')\!+\!\partial_\ell\mathfrak{m}^*_{12}(\omega,\!\omega')\mathfrak{m}_{12}(\omega,\!\omega') \right]\;.
\end{eqnarray} 
	
\begin{figure}[!t]
		\centering
		\resizebox{0.8\textwidth}{!}{\includegraphics{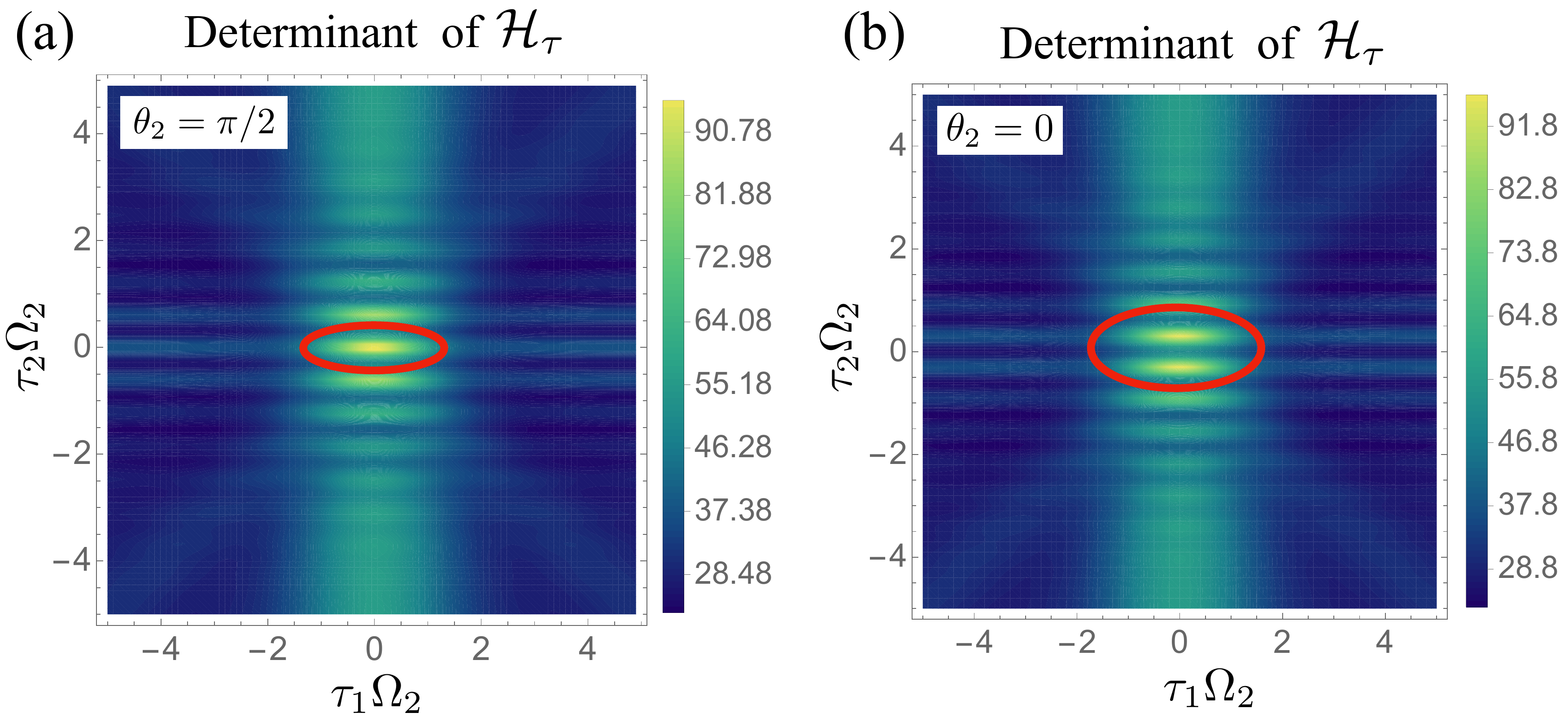}}
		\caption{In panel (a) we report the determinant of the QFI matrix for two time-delay parameters, equation~(\ref{eq:completeQFI}), computed for the symmetric biphoton pure state having JSA expressed by  equation~(\ref{spectrum}), with the achromatic phase $\theta_2=\pi/2$ ($\theta_2$ is involved in the phase-shift module $\varphi_2(\omega)$ of the GHOM setting as shown in figure~\ref{Fig:GHOM} and equation~(\ref{eq:phase_shift})). Panel (b) represents the same quantity but in absence of the {achromatic} phase, i.e. $\theta_2=0$. In both cases the determinant is non-null for the plotted region of parameters. Red ellipses mark the regions where the determinants assume their maximal values. Here $\tau_1$ and $\tau_2$ are always rescaled by the inverse of the width $\Omega_2$ of the biphoton JSA function (see  equation~(\ref{spectrum})), and $\Omega_1=\Omega_2/3$, $\omega_0=5\Omega_2,{\Omega_2=1}$ are set for the simulation.}
		\label{fig:jointDet} 
	\end{figure}

\begin{figure}[!t]
		\centering
		\resizebox{1.0\textwidth}{!}{\includegraphics{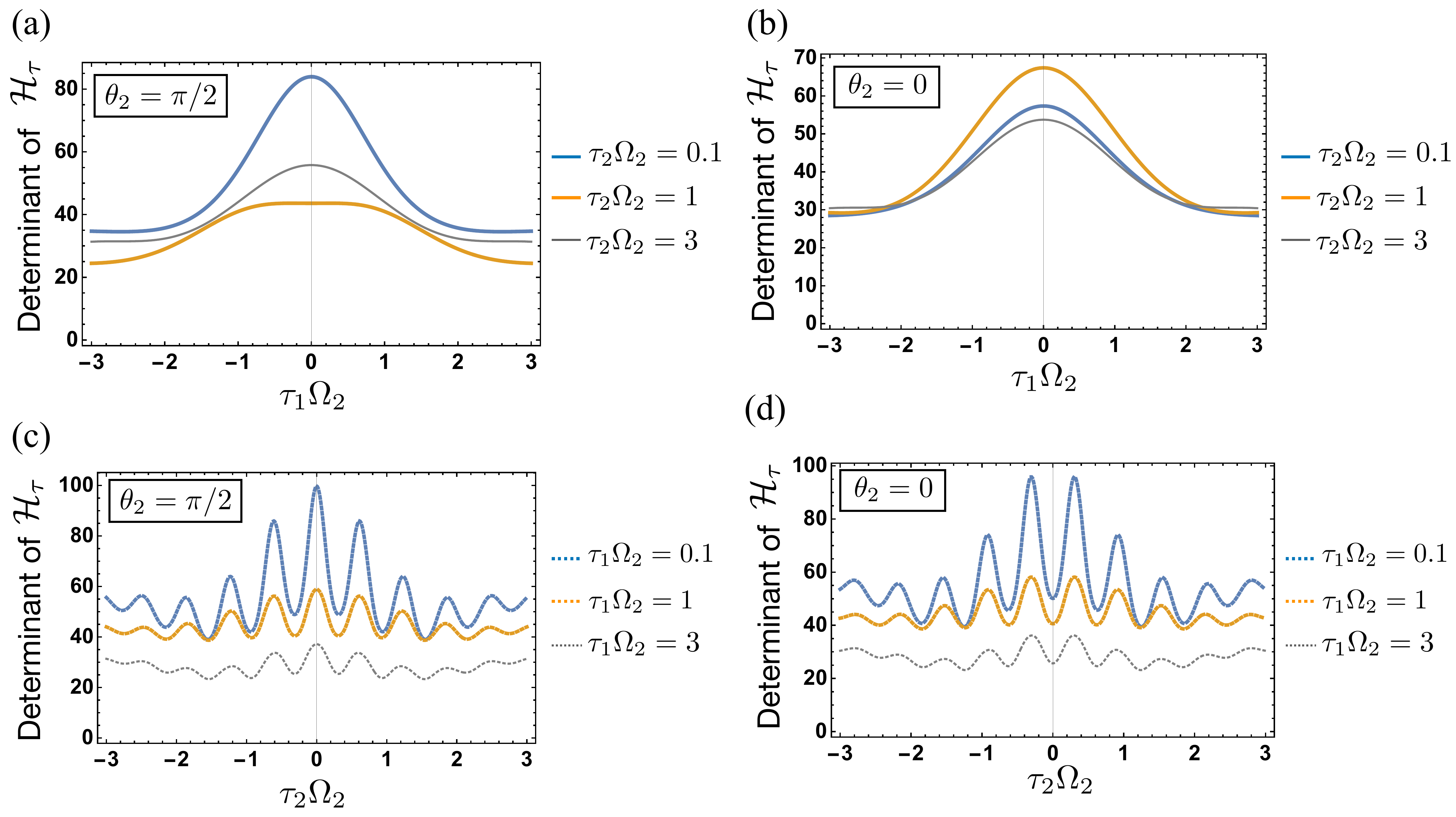}}
		\caption{In all the above panels the determinant of the QFI {matrix} in equation~(\ref{eq:completeQFI}) is plotted, computed for the symmetric biphoton pure state having JSA expressed by equation~(\ref{spectrum}). Panel (a) and (b) show the determinant as a function of $\tau_1$ respectively for {$\theta_2 = \pi/2$ and $\theta_2 = 0$}, while panel (c) and (d) show the determinant as a function of $\tau_2$ again respectively for {$\theta_2 = \pi/2$ and $\theta_2 = 0$}. In the plotted region of parameters the determinant is non-null and the QFI matrix is invertible. Here $\tau_1$ and $\tau_2$ are always rescaled by the inverse of the width $\Omega_2$ of the biphoton JSA function (see equation~(\ref{spectrum})), and $\Omega_1=\Omega_2/3$, $\omega_0=5\Omega_2,{\Omega_2=1}$ are set for the simulation.}
		\label{fig:jointExtra} 
	\end{figure}

\begin{figure}[!t]
		\centering
		\resizebox{1.1\textwidth}{!}{\includegraphics{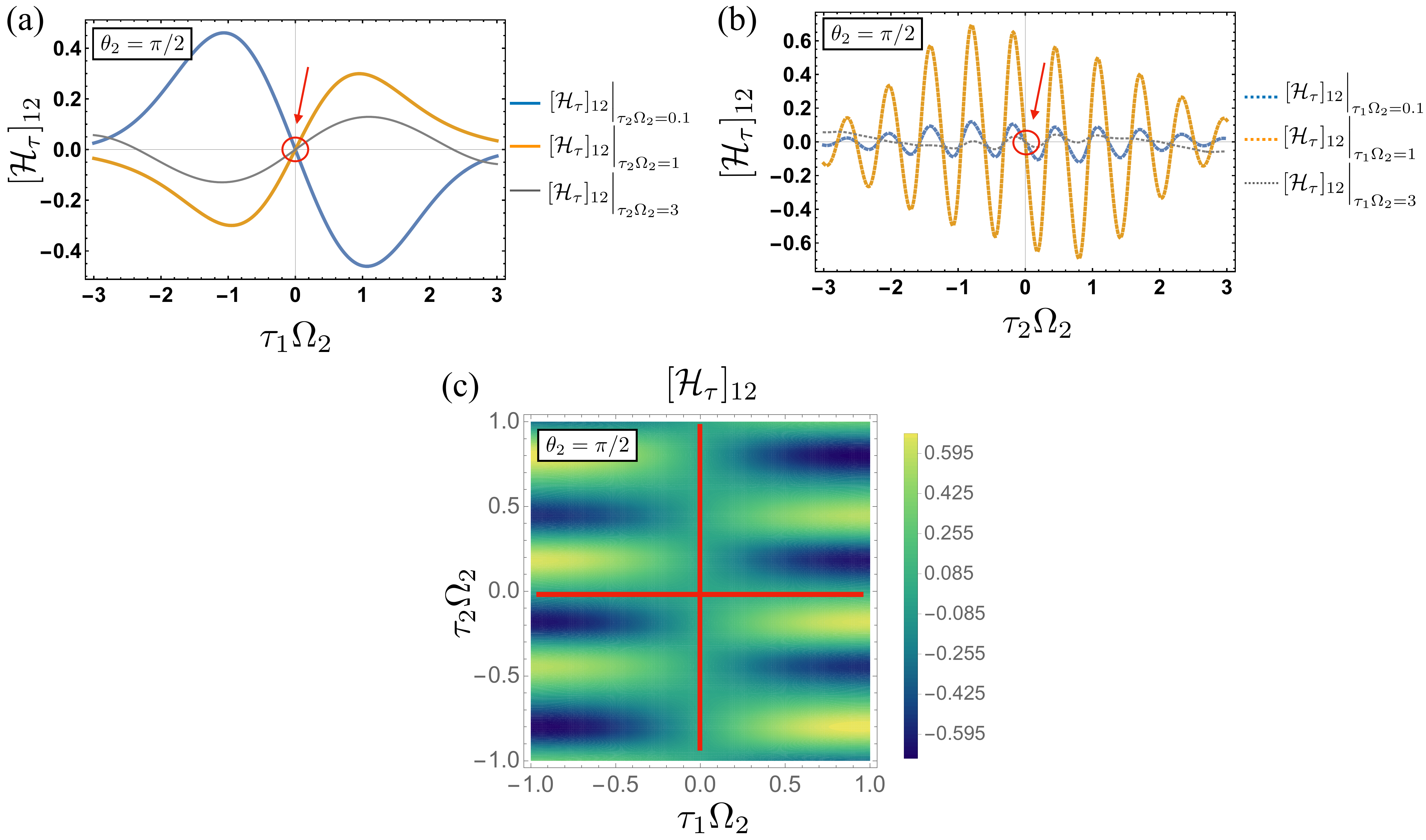}}
		\caption{Panels (a) and (b) respectively give the functional dependence upon $\tau_1$ and $\tau_2$ of the off-diagonal term $[\bm{\mathcal{H}_{\tau}}]_{12}|_{\theta_2=\pi/2}$ for the JSA function equation~(\ref{spectrum}) as computed in equation~(\ref{DEFH12}) of~\ref{APPEVALUES}. Countourplot (c) shows the functional dependence of $[\bm{\mathcal{H}_{\tau}}]_{12}|_{\theta_2=\pi/2}$ of equation~(\ref{DEFH12}) on the full $(\tau_1,\tau_2)$ plane, in which the zero values of $[\bm{\mathcal{H}_{\tau}}]_{12}|_{\theta_2=\pi/2}$ locate along two perpendicular red lines representing $\tau_1=0$ and $\tau_2=0$ in agreement with equation~(\ref{ZERO}). Here $\tau_1$ and $\tau_2$ are rescaled by the inverse of the width $\Omega_2$ of the biphoton JSA function (see  equation~(\ref{spectrum})), and $\Omega_1=\Omega_2/3$, $\omega_0=5\Omega_2,\Omega_2=1,\theta_2=\pi/2$ are set for the simulation.}
		\label{figH122} 
	\end{figure}

To get further insight on the structure of the matrix $\bm{\mathcal{H}_{\tau}}$ we now take a paradigmatic two-mode Gaussian function to be the JSA $\Psi_s(\omega,\omega')$ of the input signal, i,e.
\begin{eqnarray}\label{spectrum}
|\Psi_s(\omega,\omega')|^2=\frac{1}{\sqrt{2\pi } \Omega_1 }e^{-\frac{(\omega+\omega'-2\omega_0)^2}{8\Omega_1^2}}  \times \frac{1}{\sqrt{2\pi}\Omega_2} e^{-\frac{(\omega-\omega')^2}{2\Omega_2^2}},
\end{eqnarray}
which locally allocates to each photon an average frequency $\omega_0$ and a spread $\Delta \omega=\sqrt{\Omega_2^2+4\Omega_1^2}/2$. Inserting equation~(\ref{spectrum}) into equation~(\ref{FINALFORQ}) all the integrals present in such expressions can be analytically solved allowing for close, yet cumbersome, formulas which we report in \ref{APPEVALUES}. The resulting function allows for an explicit evaluation of the determinant of the QFIM that turns out to be always strictly positive as shown in   figures~\ref{fig:jointDet} and~\ref{fig:jointExtra}. Here $\mbox{Det}[\bm{\mathcal{H}_{\tau}}]$ is plotted as a function of $\tau_1$ and $\tau_2$ for $\theta_2 = \overline{\theta}_2= \pi/2$ (special choice of $\theta_2$ that enable the EZC condition~(\ref{relation})), and  for  $\theta_2 = 0$ (no achromatic phases). 
Notice that  for both these values of $\theta_2$ the $\mbox{Det}[\bm{\mathcal{H}_{\tau}}]$ is not null meaning that  to untwin $\tau_1$ and $\tau_2$ there is not need to enforce the EZC condition (the mere presence of the BSs suffice for this scope). Yet, setting the achromatic phase $\theta_2 = \pi/2$  the estimation model seems to yield some extra  advantages that appears as we {analyze} in details the various elements of the QFIM. 
	\begin{figure}[!t]
		\centering
		\resizebox{0.8\textwidth}{!}{\includegraphics{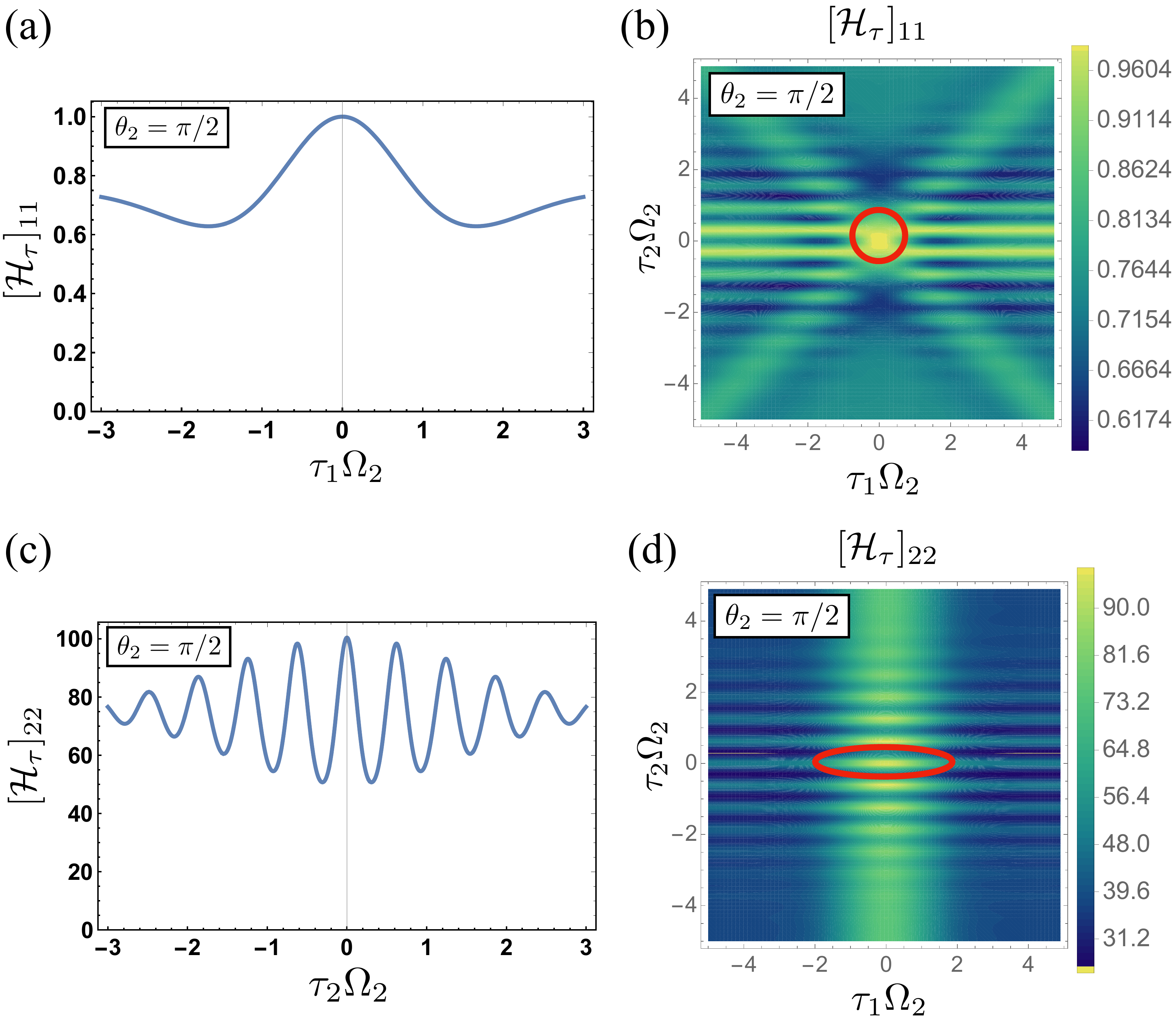}}
		\caption
		{Panel (a) gives the dependence of $\left[ \bm{\mathcal{H}_{\tau}}\right]_{11}\Big|_{\tau_2=0, \theta_2={\pi}/{2}}$ with respect to $\tau_1$ as reported in equation~(\ref{eq:QFI1new}); similarly plane (c) gives the dependence of $\left[ \bm{\mathcal{H}_{\tau}}\right]_{22}\Big|_{\tau_1=0, \theta_2={\pi}/{2}}$ with respect to $\tau_2$ as reported in equation~(\ref{eq:QFI2new}){. The} countourplots (b) and (d) give respectively the functional dependence of $\left[ \bm{\mathcal{H}_{\tau}}\right] _{11}$ and {$\left[ \bm{\mathcal{H}_{\tau}}\right] _{22}$} on the full $(\tau_1,\tau_2)$ plane determined by equations~(\ref{EXPH11}) and (\ref{EXPH22}) of~\ref{APPEVALUES} -- in both case the region where they assume their maximal values (\ref{MAX}) are marked by red ellipses. Here $\tau_1$ and $\tau_2$ are always rescaled by the inverse of the width $\Omega_2$ of the biphoton JSA function (see equation~(\ref{spectrum})), and $\Omega_1=\Omega_2/3$, $\omega_0=5\Omega_2, \Omega_2=1, \theta_2=\pi/2$ are set for the simulation.}
		\label{FigH11H22} 
	\end{figure}
	Plots of all these expressions as a function of the delay parameters ${\tau_1}$ and ${\tau_2}$ are presented in figures~\ref{FigH11H22} and \ref{figH122}. In particular the first show that when $\theta_2=\pi/2$, both $\left[ \bm{\mathcal{H}_{\tau}}\right]_{11}$ and $\left[ \bm{\mathcal{H}_{\tau}}\right]_{22}$ simultaneously exhibits a global maximum in correspondence of the EZC point $\tau_1=\tau_2=0$,
	\begin{equation} 
		\eqalign{
		\left[\bm{\mathcal{H}_{\tau}}\right]_{11}^{(\max)}\Big|_{{\theta_2}={\pi/2}} =\left[ \bm{\mathcal{H}_{\tau}}\right]_{11}\Big|_{{\bm \tau}={\bm 0},{\theta_2}={\pi/2}} =\Omega_2^2\;, \nonumber\\ \left[\bm{\mathcal{H}_{\tau}}\right]_{22}^{(\max)} \Big|_{{\theta_2}={\pi/2}} = \left[ \bm{\mathcal{H}_{\tau}}\right]_{22}\Big|_{{\bm \tau}={\bm 0},{\theta_2}={\pi/2}} =4 ( \omega_0^2 +\Omega_1^2)\;. } \label{MAX}
	\end{equation} 
	It actually turns out that irrespective on the specific choice of the input state given in (\ref{spectrum}) the off-diagonal term $\left[ \bm{\mathcal{H}_{\tau}}\right]_{12} \Big|_{\theta_2 = \pi/2}$ is null on the principal axis of the $(\tau_1, \tau_2)$ space, i.e.
	\begin{eqnarray} \label{ZERO} 
		\left[ \bm{\mathcal{H}_{\tau}}\right]_{12}\Big|_{\tau_1=0,\theta_2=\frac{\pi}{2}}=\left[ \bm{\mathcal{H}_{\tau}}\right]_{12}\Big|_{\tau_2=0,\theta_2=\frac{\pi}{2}}=0\;. 
	\end{eqnarray}
	Indeed from  equation~(\ref{eq:completeQFI}) the off-diagonal component of the QFI matrix  can be expressed as
	\begin{eqnarray}\label{FINALFORQsimp}
\hspace{-2.5cm}[\bm{\mathcal{H}_{\tau}}]_{12}\Big|_{\theta_2=\frac{\pi}{2}}&\!=\!&{-\frac{1}{2}}\int\! d\omega \int\! d\omega' |\Psi_s(\omega,\omega')|^2 \nonumber\\
&\! \times \! & \omega (\omega-\omega') \sin (\tau_2 \omega) \cos (\tau_2 \omega') \sin (\tau_1 (\omega-\omega')) + 4 \mathfrak{h}_1(\tau_1,\tau_2)\mathfrak{h}_2(\tau_1,\tau_2)\;, 
	\end{eqnarray}
	with
	\begin{eqnarray}
		\fl \mathfrak{h}_1(\tau_1,\tau_2)&=&\frac{1}{8} \int\! d\omega \int\! d\omega' |\Psi_s(\omega,\omega')|^2 (\omega-\omega') \sin (\tau_1 (\omega-\omega')) \cos (\tau_2 \omega) \cos (\tau_2 \omega')\;, \\ 
		\fl \mathfrak{h}_2(\tau_1,\tau_2)&=&\frac{1}{4} \int\! d\omega \int\! d\omega' |\Psi_s(\omega,\omega')|^2 \sin (\tau_2 \omega) \cos (\tau_2 \omega') (\omega \cos (\tau_1 (\omega-\omega'))+\omega') \;,
	\end{eqnarray}
	from which it is easy to verify~(\ref{ZERO}). In particular this holds at the EZC point implying that under such condition the QFIM of the problem is diagonal so that the gap between the second and the third terms of equation~(\ref{tmp1}) saturates, and the two time-delay parameters are statistically independent. Thus it can be seen that the employment of $\theta_2=\pi/2$ is not only necessary for enforcing the EZC condition, but also pivotal for achieving the statistical independence between two to-be-estimated time-delay parameters.

\subsubsection{Weak commutativity condition}\label{sec:weak}
Here we show that at the output of the GHOM interferometer the {condition~(\ref{weakcomm_new})} is met for all values of ${\bm \tau}$. Indeed exploiting the identities~(\ref{NEWIDE}), equation~(\ref{weakcomm_new}) simplifies as
\begin{eqnarray}\label{weakcomm_new_1}
		\text{Im}[ \langle \partial_1 \Phi_{\bm \tau}|\partial_2 \Phi_{\bm \tau}\rangle+ \langle \partial_1 \Upsilon_{\bm \tau}|\partial_2 \Upsilon_{\bm \tau}\rangle] =0\;.
\end{eqnarray}
To verify that such an identity is indeed valid we now notice that invoking equations~(\ref{m1new})--(\ref{m3new}) one can show that $\langle \partial_1 \Phi_{\bm \tau}|\partial_2 \Phi_{\bm \tau}\rangle$ is a real quantity, i.e. 
\begin{eqnarray} 
\hspace{-1cm}	&& \langle \partial_1 \Phi_{\bm \tau}|\partial_2 \Phi_{\bm \tau}\rangle\nonumber\\
\hspace{-1cm} &\!=\!&\int \! d\omega \! \int \! d\omega' |\Psi_s(\omega,\omega')|^2 [ \partial_1 \mathfrak{m}^*_{11}(\omega,\omega')\partial_2 \mathfrak{m}_{11}(\omega,\omega')\!+\!\partial_1\mathfrak{m}^*_{22}(\omega,\omega')\partial_2\mathfrak{m}_{22}(\omega,\omega')] \nonumber \\
\hspace{-1cm}	&=& \frac{1}{32} \int \! d\omega \! \int \! d\omega' |\Psi_s(\omega,\omega')|^2 \sin (\tau_1 (\omega-\omega'))\nonumber\\ 
\hspace{-1cm} &\times&[(\omega-\omega')^2 \sin (\tau_2 (\omega-\omega'))-(\omega^2-{\omega'}^{2}) \sin (2 \theta_2 +\tau_2 (\omega+\omega'))] \nonumber \\ 
\hspace{-1cm}	&=& \langle \partial_2 \Phi_{\bm \tau}|\partial_1 \Phi_{\bm \tau}\rangle \;, 
\end{eqnarray}
and that 
\begin{eqnarray} 
\fl		&&	{\langle \partial_1 \Upsilon_{\bm \tau}|\partial_2 \Upsilon_{\bm \tau}\rangle-\langle \partial_2 \Upsilon_{\bm \tau}|\partial_1 \Upsilon_{\bm \tau}\rangle}\nonumber\\
\fl		&=& \int \! d\omega \! \int \! d\omega' |\Psi_s(\omega,\omega')|^2 
		{[\partial_1\mathfrak{m}^*_{12}(\omega,\omega')\partial_2 \mathfrak{m}_{12}(\omega,\omega')-\partial_2 \mathfrak{m}^*_{12}(\omega,\omega')\partial_1 \mathfrak{m}_{12}(\omega,\omega')]}\nonumber \\ 
\fl		&=&\frac{i}{8} \int \! d\omega \! \int \! d\omega' |\Psi_s(\omega,\omega')|^2 (\omega-\omega') e^{-i \tau_1 (\omega+\omega')} \Big[\left(e^{2 i \tau_1 \omega}+e^{2 i \tau_1 \omega'}\right)\nonumber\\
\fl &\times& (\omega' \sin (\theta_2 +\tau_2 \omega)\!+\!\omega \sin (\theta_2 +\tau_2 \omega'))
\!+\! 2 e^{i \tau_1 (\omega+\omega')} (\omega \sin (\theta_2\!+\!\tau_2 \omega)+\omega' \sin (\theta_2 \!+\! \tau_2 \omega'))\Big] \nonumber\\
\fl &=&0\;,
\end{eqnarray}
due to the fact that the integrand is explicitly anti-symmetric for {the} exchange of $\omega$ with $\omega'$. We can hence conclude that in present case the bound~(\ref{Multi_QCRB}) saturates at least asymptotically in the limit of large $M$. Notice also that this result is valid not just when we met the EZC point condition~(\ref{relation}), but for all choices of $\tau_1$, $\tau_2$ and $\theta_2$ as long as the JSA function $|\Psi_s(\omega,\omega')|^2$ is symmetric.

\subsubsection{GHOM settings of $k=3$ and $k=4$ }\label{sec:higher}
\begin{figure*}[!h]
	\centering
	\includegraphics[width=1.05\textwidth]{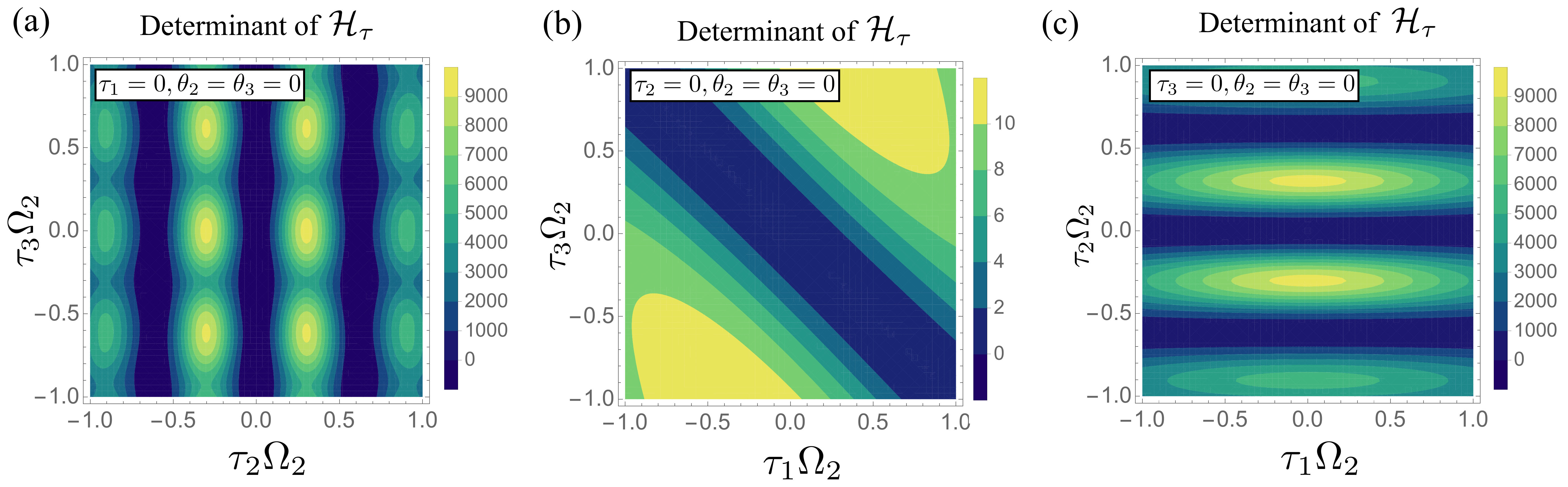}
	\caption{ Contourplots (a), (b) and (c)  respectively give  the functional dependence of the determination of the QFIM with respect to  three time delays upon the full  $(\tau_2,\tau_3)$, $(\tau_1,\tau_3)$ and $(\tau_1,\tau_2)$ planes with the given value $\tau_1=0$, $\tau_2=0$ and $\tau_3=0$.  The achromatic phase shifts $\theta_2$ and $\theta_3$ are set to be zeros for simplifying the calculation. Here $\tau_1$, $\tau_2$ and $\tau_3$ are rescaled by the inverse of the width $\Omega_2$ of the biphoton JSA function (see  equation~(\ref{spectrum})), and $\Omega_1=\Omega_2/3$, $\omega_0=5\Omega_2$, $\Omega_2=1$ are set for the simulation.}
	\label{Det3D}
\end{figure*}

\begin{figure*}[h]
	\centering
	\includegraphics[width=1.05\textwidth]{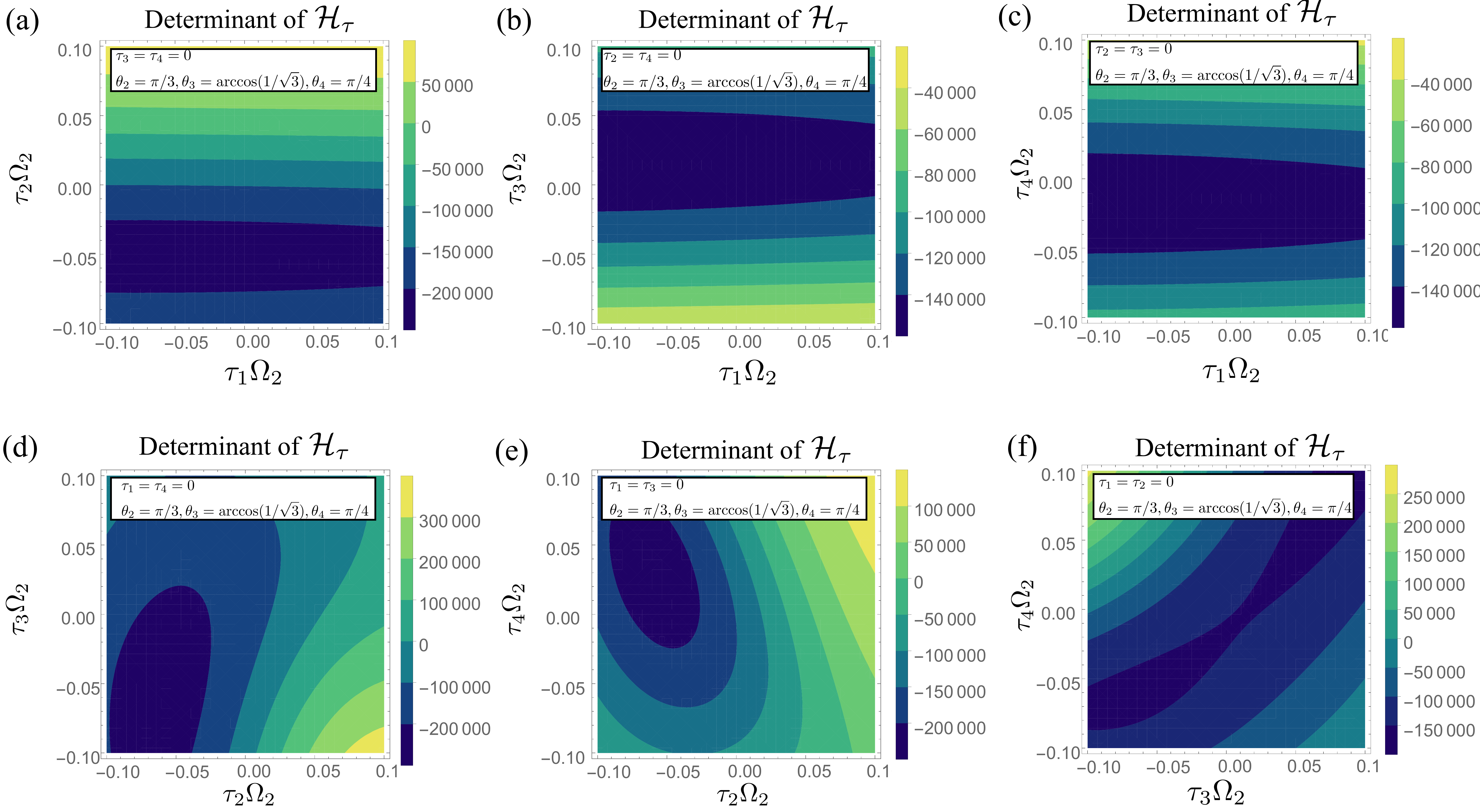}
	\caption{Contourplots (a)-(f) give respectively the functional dependence of the determination of the QFIM with respect to  four time delays upon the full  $(\tau_1,\tau_2)$, $(\tau_1,\tau_3)$, $(\tau_1,\tau_4)$,  $(\tau_2,\tau_3)$, $(\tau_2,\tau_4)$ and $(\tau_3,\tau_4)$ planes, in which $\theta_2=\pi/3$, $\theta_4=\pi/4$, $\theta_3=\arccos\left({1}/{\sqrt{3}}\right)$ and the corresponding remaining parameters are set to be zeros. Here $\tau_1$, $\tau_2$, $\tau_3$ and $\tau_4$ are rescaled by the inverse of the width $\Omega_2$ of the biphoton JSA function (see  equation~(\ref{spectrum})), and $\Omega_1=\Omega_2/3$, $\omega_0=5\Omega_2$, $\Omega_2=1$ are set for the simulation.}
	\label{Det4D}
\end{figure*}

In this section, we focus on the GHOM estimation scheme with a higher $k$, for instance $k=3$ and $4$, and try to explore whether the invertibility of the QFIM still keeps under the EZC condition.
Firstly to analyze the singularity of the QFIM with respect to three time-delay parameters on the point $\tau_1=\tau_2=\tau_3=0$, we study the functional dependence of its determinant upon the full  $(\tau_2,\tau_3)$, $(\tau_1,\tau_3)$ and $(\tau_1,\tau_2)$ planes with the given value $\tau_1=0$, $\tau_2=0$ and $\tau_3=0$ respectively.
The simulation results  are depicted in figure~\ref{Det3D}, in which  achromatic phase shifts $\theta_2$ and $\theta_3$ are set to be zeros for simplifying the calculation (this is acceptable since whatever the values of $\theta_2,\theta_3$, the EZC condition of equation~(\ref{relation}) does not hold in the case of $k=3$~\cite{GHOM}).
Figure~\ref{Det3D} shows that the determination of the QFIM for three time-delay parameters is always zero on the point $\tau_1=\tau_2=\tau_3=0$,  which is opposite to the case of $k=2$ shown in figure~\ref{fig:jointDet}.
In other words, the QFIM is non-invertible or singular for the case of $k=3$.
If we further generalize the above consideration into the GHOM scheme with four time delays, the QFIM will be nonsingular similar to the case of $k=2$. 
To analyze the singularity of the QFIM with respect to four time-delay parameters on the point $\tau_1=\tau_2=\tau_3=\tau_4=0$, we respectively study the functional dependence of its determinant upon the full  $(\tau_1,\tau_2)$, $(\tau_1,\tau_3)$, $(\tau_1,\tau_4)$,  $(\tau_2,\tau_3)$, $(\tau_2,\tau_4)$ and $(\tau_3,\tau_4)$ planes with the remaining parameters are set to be zeros.
The simulation results are exhibited in figure~\ref{Det4D} where the corresponding achromatic phase shifts $\theta_2=\pi/3, \theta_3=\arccos\left(\frac{1}{\sqrt{3}}\right), \theta_4=\pi/4$ (this configuration makes the EZC condition of equation~(\ref{relation}) hold in the case of $k=4$~\cite{GHOM}).

\section{Conclusions}\label{Sec:Conclusion}
In the present work we have presented a method that untwines multiple parameters from an intertwined multi-parameter scenario and achieves the multi-delays simultaneous optimal estimation by introducing some necessary quantum controls. As an explicit example, we have showed the untwining of two time-delay parameters in a GHOM interferometer, and proved that the resulting estimation scheme is in many ways optimal around the EZC point. Specifically, at the EZC point every time-delay parameter can be estimated with the individual highest precision (the diagonal elements of the QFIM reach the individual maxima) and the statistical independence between them can be achieved (the off-diagonal elements of the QFIM are zeros). Furthermore, the multi-parameter QCRB can be saturated at least in the asymptotic limit of infinite copies (the weak commutation condition can be satisified). Finally, we prove that the similar results have been obtained in the case of  $k=4$ time delays with a GHOM interferometer under the EZC condition ${\bm{\theta}=\bm{\bar{\theta}}}$ (specifically we have numerically tested the model using $\theta_2=\pi/3, \theta_3=\arccos\left({1}/{\sqrt{3}}\right), \theta_4=\pi/4$). Notably instead for $k=3$ time delays we haven't been able to find a configuration of $\theta_i$ such that the QFIM is nonsingular for $\tau_i = 0$ ($i=1,2,3$), a fact that  mimics the observation of reference~\cite{2HOM} that no EZC point can be found under this condition.

From the perspective of realistic application, the current illustration described by the GHOM interferometry with a set of unknown time delays is  one of  representative problems of multi-phase estimation, which could inspire many applications like developing the monitoring of terrain deformation or the photogrammetry by virtue of the Interferometric Synthetic Aperture Radar (InSAR) techniques~\cite{InSAR1,InSAR2}.

\ack
This work is supported by the Shaanxi Natural Science Basic Research Program (Grant No. 2021JQ-008); National Natural Science Foundation of China (NSFC) (Grants No. 12204371, 12074307 and 62071363).
F.B. and V.G.  acknowledge MIUR  (Ministero dell' istruzione, dell' Universita' e della Ricerca) via project PRIN 2017 ``Taming complexity via Quantum Strategies a Hybrid Integrated Photonic approach" (QUSHIP) Id. 2017SRNBRK.

\appendix
	\section{Two-photon output state emerging from a $k=2$ GHOM interferometer}\label{app:state}
	The terms which define the output state (\ref{out})  emerging  from the GHOM interferometer can be formally written as
	\begin{equation}\label{coin1}
	\fl\eqalign{
	|\Phi_{\bm \tau}\rangle=\int d\omega \int d\omega' \Psi_s(\omega,\omega') [ \mathfrak{m}_{11}(\omega,\omega') \hat{c}_1^{\dagger}(\omega)\hat{c}_1^{\dagger}(\omega') + \mathfrak{m}_{22}(\omega,\omega') \hat{c}_2^{\dagger}(\omega)\hat{c}_2^{\dagger}(\omega')]|\varnothing\rangle\;,\\
	|\Upsilon_{\bm \tau}\rangle=\int d\omega \int d\omega' \Psi_s(\omega,\omega')\mathfrak{m}_{12}(\omega,\omega')\hat{c}_1^{\dagger}(\omega)\hat{c}_2^{\dagger}(\omega')|\varnothing\rangle\;,}
	\end{equation}
	with $\hat{c}_{1}^{\dagger}(\omega)$, $\hat{c}_{2}^{\dagger}(\omega)$ being the output Bosonic creation operators associated respectively to the upper and lower harm of the setup, and with $\mathfrak{m}_{11}(\omega,\omega')$, $\mathfrak{m}_{22}(\omega,\omega')$, $\mathfrak{m}_{12}(\omega,\omega')$ the transition amplitudes that encode the full dependence upon the parameter vector ${\bm \tau}$ and the achromatic wave-plates vector ${\bm \theta}:=\{\theta_2,\cdots, \theta_k\}$. While referring the reader to reference~\cite{GHOM} for the general case, here we focus  on the special case $k=2$ that is the subject of the present investigation. 
	In this case the linear mapping between the creation operators at the input of the device and the creation operators at the output writes as
	\begin{eqnarray}\label{eq:mapping}
	\left( \begin{array}{lcr} \hat{a}_1^{\dagger}(\omega)\\
			\hat{a}_2^{\dagger}(\omega) \end{array} \right)
		 = \mathbf{M}(\omega)
		  \left( \begin{array}{lcr} \hat{c}_1^{\dagger}(\omega)\\
		 	\hat{c}_2^{\dagger}(\omega) \end{array}\right)\;,
	\end{eqnarray}
	with the $2\times2$ transformation matrix
\begin{eqnarray}
\mathbf{M}(\omega) \!=\!\left( \begin{array}{lcr} e^{i\frac{\omega \tau_1}{2}}\cos(\frac{\omega \tau_2+\theta_2}{2}) & ie^{i\frac{\omega \tau_1}{2}}\sin(\frac{\omega \tau_2+\theta_2}{2}) \\
ie^{-i\frac{\omega \tau_1}{2}}\sin(\frac{\omega \tau_2+\theta_2}{2}) & e^{-i\frac{\omega \tau_1}{2}}\cos(\frac{\omega \tau_2+\theta_2}{2})  \end{array} \right).\label{Newmap2}
\end{eqnarray}
Expressing $\hat{a}_1^{\dagger}(\omega)$ and $\hat{a}_2^{\dagger}(\omega)$ in terms of $\hat{c}_1^{\dagger}(\omega)$ and $\hat{c}_2^{\dagger}(\omega)$ via (\ref{eq:mapping}), and replacing the result into equation~(\ref{eq:input})  yields a final state of the system~(\ref{out})  with the vectors $|\Phi_{\bm \tau}\rangle$ and $|\Upsilon_{\bm \tau}\rangle$ of  equation~(\ref{coin1}) defined by the amplitudes
\begin{eqnarray}
\fl		\mathfrak{m}_{11}(\omega,\omega')&:=&\frac{-i}{4}e^{-i\left( \frac{\tau_1(\omega\!+\! \omega')}{2}\right) } \nonumber\\
\fl &\times& \!\left[( e^{i\tau_1 \omega}\!-\! e^{i\tau_1 \omega'}) \sin \! \left(\frac{\tau_2(\omega\!-\!\omega')}{2} \! \right) \!-\! ( e^{i\tau_1 \omega}\!+\!e^{i\tau_1 \omega'}) \sin \! \left(\frac{\tau_2(\omega+\omega')}{2}\!+\!\theta_2 \! \right)\! \right],\label{m1new}\\ 
\fl \mathfrak{m}_{22}(\omega,\omega')&:=&\frac{i}{4}e^{-i\left(\frac{\tau_1(\omega\!+\! \omega')}{2} \right) }\nonumber\\ 
\fl &\times&\!\left[(e^{i\tau_1\omega}\!-\! e^{i\tau_1\omega'} ) \sin \!\left(\frac{\tau_2(\omega\! -\! \omega')}{2} \!\right) \!+\! (e^{i\tau_1\omega}\!+\! e^{i\tau_1\omega'} ) \sin \!\left(\frac{\tau_2(\omega\!+\! \omega')}{2} +\theta_2\! \right)\! \right],\label{m2new}\\ 
\fl	\mathfrak{m}_{12}(\omega,\omega') &:=&\frac{1}{2}e^{-i\left( \frac{\tau_1(\omega\!+\!\omega')}{2}\right) }\nonumber\\
\fl &\times& \!\left[(e^{i\tau_1 \omega}\!-\! e^{i\tau_1 \omega'} )\cos \! \left( \frac{\tau_2(\omega\!-\! \omega')}{2}\! \right) \!+\! (e^{i\tau_1 \omega}\!+\! e^{i\tau_1 \omega'} ) \cos\! \left(\! \frac{\tau_2 (\omega\!+\! \omega')}{2} \!+\! \theta_2 \! \right) \! \right]\;. \label{m3new}
\end{eqnarray}
For the sake of completeness we observe that for an input state with JSA spectrum given in equation~(\ref{spectrum}) the coincidence count  probability~(\ref{RBP}) becomes 
\begin{eqnarray}\label{eq:pro}
R(\bm{\tau})&=&\frac{1}{8} \left[4 - e^{ \frac{-(\tau_1-\tau_2)^2 \Omega_2^2}{2} }+2e^{\frac{-\tau_2^2 \Omega_2^2}{2} } -e^{ \frac{-(\tau_1+\tau_2)^2 \Omega_2^2}{2}} \right. \nonumber\\
&+&\left. 2e^{-2\tau_2^2 \Omega_1^2-\frac{1}{2}\tau_1^2 \Omega_2^2}\left( 1+e^{\frac{\tau_1^2 \Omega_2^2}{2}}\right) \cos\left(2\tau_2 \omega_0+ 2\theta_2 \right) \right],
\end{eqnarray}
which for $\theta_2=\pi/2$ can be easily seen to meet the EZC condition~(\ref{relation})~\cite{GHOM}. 

\section{Values of the QFIM entries for a Gaussian JSA spectrum} \label{APPEVALUES}
Here we report the explicit values of the QFIM entries associated with an input state with the JSA spectrum given in equation~(\ref{spectrum}) and used to produce the plots in the main text. For diagonal entries we get
\begin{eqnarray}
&& \fl  \left[\bm{\mathcal{H}_{\tau}}\right]_{11}\Big|_{\theta_2=\frac{\pi}{2}}=-\frac{\Omega_2^4}{256}\left[e^{-\frac{(\tau_1+\tau_2)^2 \Omega_2^2}{2}} \left(\tau_1+\tau_2+e^{2\tau_1 \tau_2 \Omega_2^2}(\tau_1-\tau_2)\right) \right.\nonumber\\
&+&\left.2e^{-2\tau_2^2 \Omega_1^2-\frac{\tau_1^2 \Omega_2^2}{2}} \tau_1 \cos(2\tau_2 \omega_0)\right]^2\!+\!
\frac{\Omega_2^2}{16}\left[12+e^{-\frac{(\tau_1-\tau_2)^2 \Omega_2^2}{2}}\left(1\!-\!(\tau_1-\tau_2)^2 \Omega_2^2\right)\right.\nonumber\\ &+&\left.e^{-\frac{(\tau_1+\tau_2)^2\Omega_2^2}{2}}\left(1-(\tau_1+\tau_2)^2\Omega_2^2+2e^{\frac{\tau_1(\tau_1+2\tau_2)\Omega_2^2}{2}}(1-\tau_2^2 \Omega_2^2)\right)\right.\nonumber\\ 
&-&\left.2e^{-2\tau_2^2 \Omega_1^2-\frac{\tau_1^2 \Omega_2^2}{2}}\left(e^{\frac{\tau_1^2 \Omega_2^2}{2}}+\tau_1^2 \Omega_2^2-1\right)\cos(2\tau_2 \omega_0)\right]\;, \label{EXPH11} \\ 
&& \fl \left[\bm{\mathcal{H}_{\tau}}\right]_{22}\Big|_{\theta_2=\frac{\pi}{2}}=-\frac{e^{-4\tau_2^2\Omega_1^2-\tau_1^2\Omega_2^2-(\tau_1+\tau_2)^2\Omega_2^2}}{256}\left[e^{2\tau_2^2\Omega_1^2+\frac{\tau_1^2\Omega_2^2}{2}}\right.\nonumber\\
 &\times&\left.\left(\tau_1-e^{2\tau_1 \tau_2 \Omega_2^2}\tau_1+\tau_2+e^{2\tau_1\tau_2 \Omega_2^2}\tau_2-2e^{\frac{1}{2}\tau_1(\tau_1+2\tau_2)\Omega_2^2}\tau_2\right)\Omega_2^2\right.\nonumber\\ 
 &-&\left.4 e^{\frac{(\tau_1+\tau_2)^2}{2}\Omega_2^2} \left(1+e^{\frac{\tau_1^2 \Omega_2^2}{2}}\right) \left(-2\tau_2 \Omega_1^2\cos(2\tau_2 \omega_0) +\omega_0 \sin(2\tau_2\omega_0)\right) \right]^2\nonumber\\ 
 &+&\frac{1}{16}\Big[e^{-\frac{(\tau_1+\tau_2)^2 \Omega_2^2}{2}} \left( \Omega_2^2-(\tau_1+\tau_2)^2 \Omega_2^4-e^{2\tau_1\tau_2 \Omega_2^2}\Omega_2^2 \left((\tau_1-\tau_2)^2\Omega_2^2-1\right)\right.\nonumber\\ &+&\left.6e^{\frac{(\tau_1+\tau_2)^2\Omega_2^2}{2}} (4\omega_0^2+4\Omega_1^2+\Omega_2^2)
 +2e^{\frac{\tau_1(\tau_1+2\tau_2)\Omega_2^2}{2}} \Omega_2^2 (\tau_2^2\Omega_2^2-1)\right.\nonumber\\ &+&\left.6e^{\frac{\tau_2(2\tau_1+\tau_2)\Omega_2^2}{2}}(4\omega_0^2+4\Omega_1^2-\Omega_2^2+\tau_1^2 \Omega_2^4)\right)
\!-\!8e^{-2\tau_2^2 \Omega_1^2-\frac{\tau_1^2 \Omega_2^2}{2}}
\!\Big(\!1+e^{\frac{\tau_1^2 \Omega_2^2}{2}}\! \Big)\! \nonumber\\
 &\times&\left((4\tau_2^2 \Omega_1^4-\omega_0^2-\Omega_1^2)\cos(2\tau_2\omega_0)+4\tau_2 \omega_0 \Omega_1^2 \sin(2\tau_2 \omega_0)\right)\Big]\;,\label{EXPH22}
\end{eqnarray}
while for the off-diagonal element we have 
\begin{eqnarray} \label{DEFH12}  
&& \fl \left[ \bm{\mathcal{H}_{\tau}}\right]_{12}\Big|_{\theta_2=\frac{\pi}{2}}=\frac{\Omega_2^2}{256}\left\lbrace  -64e^{-2\tau_2^2 \Omega_1^2-\frac{\tau_1^2 \Omega_2^2}{2}}\tau_1 \left(2\tau_2 \Omega_1^2\cos(2\tau_2 \omega_0)+\omega_0 \sin(2\tau_2 \omega_0)\right) \right. \nonumber\\
&+&\left.e^{-2\tau_2^2 \Omega_1^2-\frac{(\tau_1+\tau_2)^2 \Omega_2^2}{2}}
\Big[ e^{-\frac{(\tau_1+\tau_2)^2 \Omega_2^2}{2}}\left(\tau_1+\tau_2+e^{2\tau_1 \tau_2 \Omega_2^2}(\tau_1-\tau_2)\right)\right.\nonumber\\
 &+&\left.2e^{-2\tau_2^2 \Omega_1^2-\frac{\tau_1^2 \Omega_2^2}{2}}\tau_1\cos(2\tau_2 \omega_0)\Big]\right.\nonumber\\ 
\fl &\times&\left. \left[e^{2\tau_2^2\Omega_1^2}\left(\left(1-e^{2\tau_1\tau_2\Omega_2^2}\right)\tau_1+\left(1+e^{2\tau_1\tau_2\Omega_2^2}-2e^{\frac{\tau_1(\tau_1+2\tau_2)\Omega_2^2}{2}}\right)\tau_2\right)\Omega_2^2\right.\right.\nonumber\\ 
\fl&+&\left.\left.4\left(e^{\frac{(\tau_1+\tau_2)^2\Omega_2^2}{2}}+e^{\frac{\tau_2(2\tau_1+\tau_2)\Omega_2^2}{2}}\right)\left(2\tau_2\Omega_1^2\cos(2\tau_2\omega_0)+\omega_0\sin(2\tau_2\omega_0)\right)\right]\right.\nonumber\\ \fl&+&{\hspace{-0.2cm}}\left.32e^{-\frac{(\tau_1^2+\tau_2^2)\Omega_2^2}{2}} \!\left[\! -2\tau_1\tau_2 \Omega_2^2 \cosh(\tau_1\tau_2\Omega_2^2)\!+\! ((\tau_1^2+\tau_2^2)\Omega_2^2\!-\!1)\sinh(\tau_1\tau_2\Omega_2^2) \! \right]\!  \right\rbrace.  \nonumber\\
\end{eqnarray}
We report also the entries of $\bm{\mathcal{H}_{\tau}}$ evaluated along the principal axis of the $(\tau_1,\tau_2)$ space: 
\begin{eqnarray} 
\fl		\left[ \bm{\mathcal{H}_{\tau}}\right]_{11}\Big|_{\tau_2=0, \theta_2=\frac{\pi}{2}}&=&\frac{1}{16}\Omega_2^2 \left[ 12-e^{-\tau_1^2\Omega_2^2}\tau_1^2 \Omega_2^2 +4e^{-\frac{\tau_1^2 \Omega_2^2}{2}}(1-\tau_1^2 \Omega_2^2)\right]\;,\label{eq:QFI1new}\\
\fl		 \left[ \bm{\mathcal{H}_{\tau}}\right]_{22}\Big|_{\tau_1=0, \theta_2=\frac{\pi}{2}}&=&\frac{1}{8}e^{-4\tau_2^2 \Omega_1^2}\left\{-\omega_0^2\!-\!4\tau_2^2 \Omega_1^4\!+\!24e^{4\tau_2^2 \Omega_1^2}(\omega_0^2+\Omega_1^2)\!+\!\omega_0^2\cos(4\tau_2 \omega_0) \right.\nonumber\\ 
\fl &\!-\!& \left. 4\tau_2^2\Omega_1^4 \cos(4\tau_2 \omega_0)+8e^{2\tau_2^2 \Omega_1^2}\left[(\omega_0^2\!+\!\Omega_1^2\!-\!4\tau_2^2 \Omega_1^4)\cos(2\tau_2 \omega_0)\right.\right.\nonumber\\
&\!-\!&\left.\left.4\tau_2 \omega_0 \Omega_1^2 \sin(2\tau_2 \omega_0)-4\tau_2 \omega_0 \Omega_1^2 \sin(4\tau_2 \omega_0)\right]\right\}\;,\label{eq:QFI2new}
\end{eqnarray}
and $\left[ \bm{\mathcal{H}_{\tau}}\right]_{12}\Big|_{\tau_1=0, \theta_2=\frac{\pi}{2}} = \left[ \bm{\mathcal{H}_{\tau}}\right]_{12}\Big|_{\tau_2=0, \theta_2=\frac{\pi}{2}} = 0$, meaning that the parameters are statistically independent when at least one time-delay is null.

\end{document}